\documentclass[prl,twocolumn,showpacs,aps]{revtex4-1}
\usepackage{bm}
\usepackage{graphicx}
\usepackage{amssymb}
\usepackage{amsmath}
\usepackage{eufrak}
\usepackage{color}
\usepackage[utf8]{inputenc}
\usepackage[unicode=true,colorlinks=true,citecolor=blue,urlcolor=blue]{hyperref}
\usepackage{pifont}
\usepackage{ulem}

\newcommand{\nix}[1]{}

\begin{document}

\title{Optical activity of quantum wells}
\author{L.\,V.\, Kotova$^{1,2}$} 
\author{A. V. Platonov$^1$} 
\author{V. N. Kats$^1$} 
\author{V. P. Kochereshko$^1$} 
\author{S. V. Sorokin$^1$} 
\author{S. V. Ivanov$^1$} 
\author{L. E. Golub$^1$}
\affiliation{$^1$Ioffe Institute, 194021 St.~Petersburg, Russia}
\affiliation{$^2$ITMO University, 197101 St.~Petersburg, Russia}

\begin{abstract}
We report on the observation of optical activity of quantum wells 
resulting in the conversion of the light polarization state controlled by the light propagation direction.
The polarization conversion is detected in reflection measurements. We show that a pure $s$-polarized light incident on a quantum well is reflected as an elliptically polarized wave. The signal is drastically enhanced in the vicinity of the light-hole exciton resonance.
We show that the polarization conversion is caused by the spin-orbit splitting of the light hole states and the birefringence of the studied structure. The bulk inversion asymmetry constant $\beta_{h} \approx 0.14$~eV\AA \, is determined for the ground light hole subband in a 10~nm ZnSe/ZnMgSSe quantum well.
\end{abstract}

\pacs{
73.21.Fg,	
78.20.Ek,	
71.35.-y, 
42.25.Ja	
}

\maketitle{}

Studies of polarization-sensitive optical effects allow creating optical devices and give access to fundamental properties of material systems.
A very important effect intensively investigated and widely used in practice is a conversion of light polarization state~\cite{Zvezdin_Kotov,metamat}. Examples are the rotation of a linear polarization plane and the transformation of a pure linearly or circularly polarized wave into an elliptically polarized light. 
A possibility for the polarization conversion exists in systems of sufficiently low spatial symmetry. For example, birefringent media effectively rotate light polarization plane and produce light helicity. Basic examples are half- and quarter-wave plates made of birefringent crystals widely used in both laboratories and in industry.
Recently, polarization conversion has been observed in metamaterials~\cite{metamat1,metamat,swast}, twisted photonic crystal fibers~\cite{twist} and microcavities~\cite{OSHE}.
While metamaterials convert light polarization due to a special design of building blocks,
semiconductor nanostructures are birefringent as-grown.  The polarization conversion has been demonstrated in a number of experiments on quantum wells  (QWs)~\cite{birefr_g-fact,birefr_deform,birefr_local_fields,Loginov,birefr_QWs_Moldova} and quantum dots~\cite{birefr_dots,birefr_pol_conv_dots}. The low symmetry of QWs can be caused by in-plane deformations~\cite{birefr_local_fields,birefr_deform,Loginov} or by microscopic structure of interfaces~\cite{IIA_Voisin,IIA_doped}, while the birefringence of self-assembled quantum dots  appears due to their anisotropic shape~\cite{birefr_pol_conv_dots}.

\textit{Optical activity} is an effect responsible for the polarization conversion controlled by the light propagation direction.
It is present even in homogeneous systems whose point group symmetry belongs to a gyrotropic class, i.e. allows for a linear coupling between components of a vector and a pseudovector. 
Recently it has been shown that optical activity of metals is closely related to their band topology and Berry phase~\cite{Top1,Top2}.
Optical activity in the visible spectral range is useful to investigate in semiconductors where they are greatly enhanced in the vicinity of exciton resonances~\cite{AgrGinzb}.
Optical activity of bulk gyrotropic semiconductors is well established~\cite{IvchSelk_JETP,bulk_gyr}. 
QWs grown of cubic semiconductors are gyrotropic: For the growth direction (001), point symmetry group of a QW is D$_{2d}$ or C$_{2v}$ depending on a presence of the structure inversion symmetry~\cite{BeTeZnSe,D2dC2v}. 
Recent theoretical studies showed that QWs are optically active in both cases~\cite{Golub_EPL,Porubaev}.
However experimental detection of optical activity has not been  reported so far for QWs.

In this work, we address the fundamental question: whether real QWs are optically active?
We report on the observation and study of optical activity in QWs. We demonstrate a resonant enhancement of  the polarization conversion in the vicinity of the light-hole exciton transition.

\begin{figure}[t]
\includegraphics[width=0.95\linewidth]{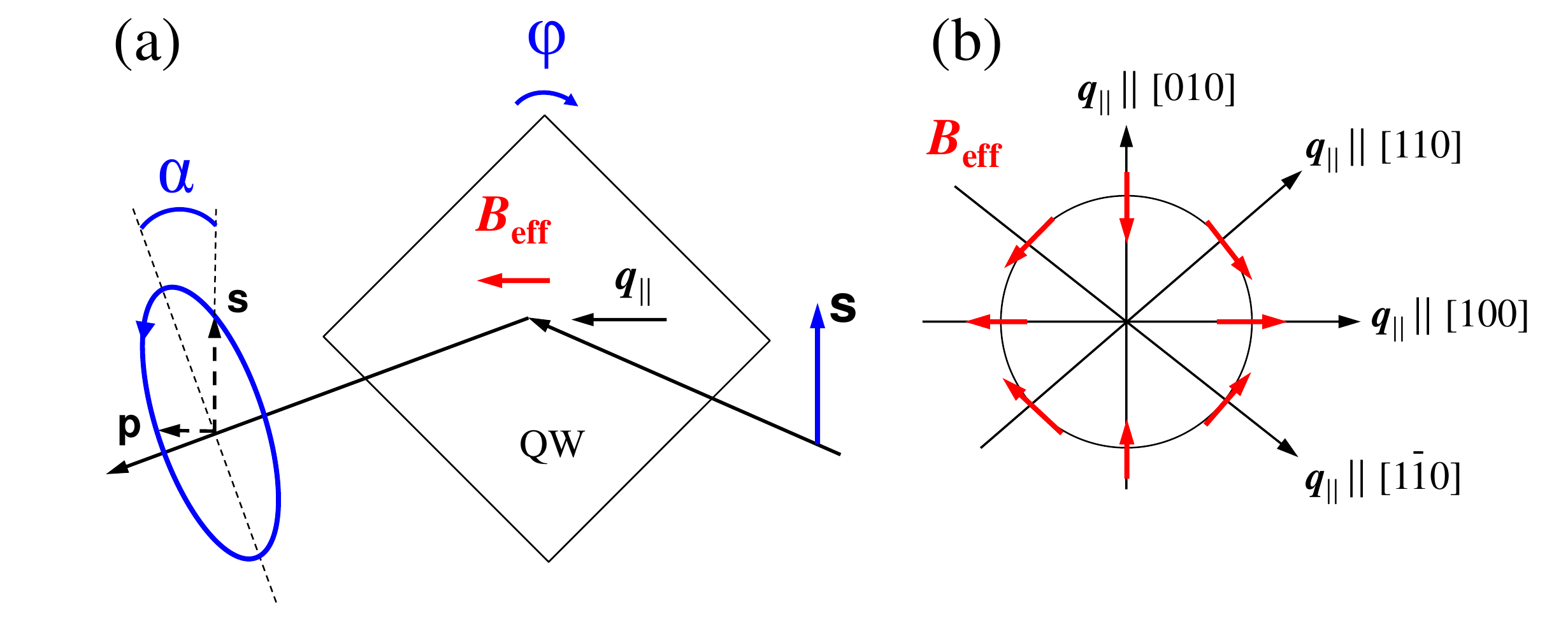}
\\
\includegraphics[width=0.45\linewidth]{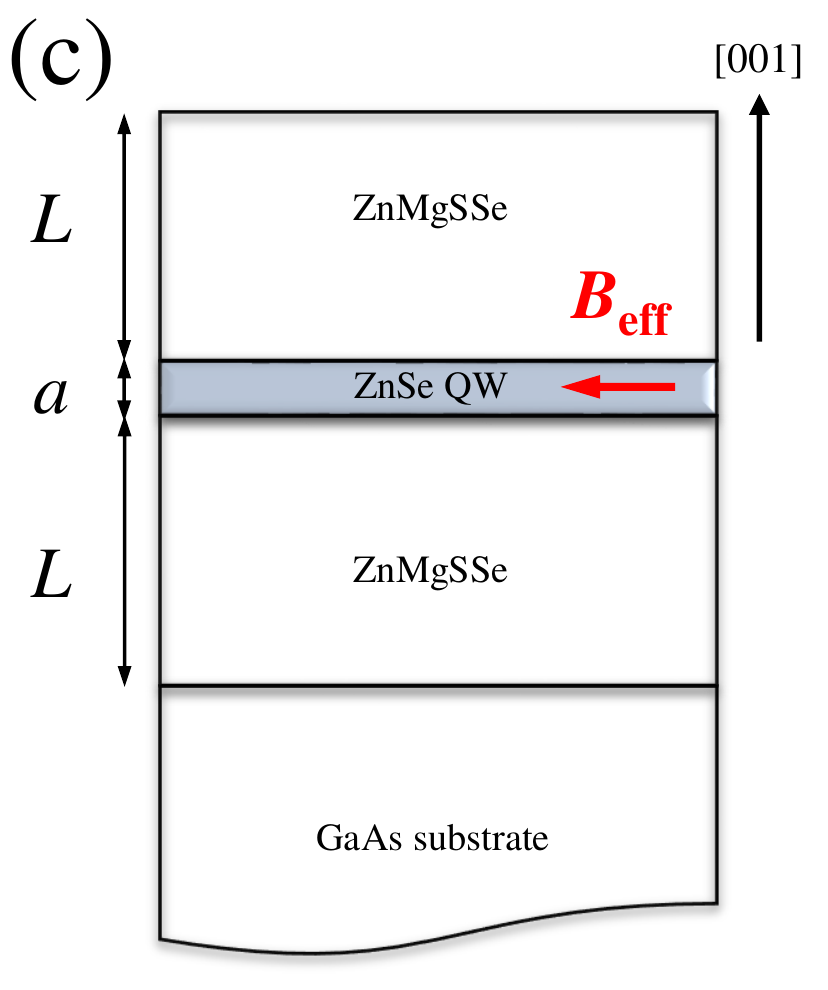}
\includegraphics[width=0.5\linewidth]{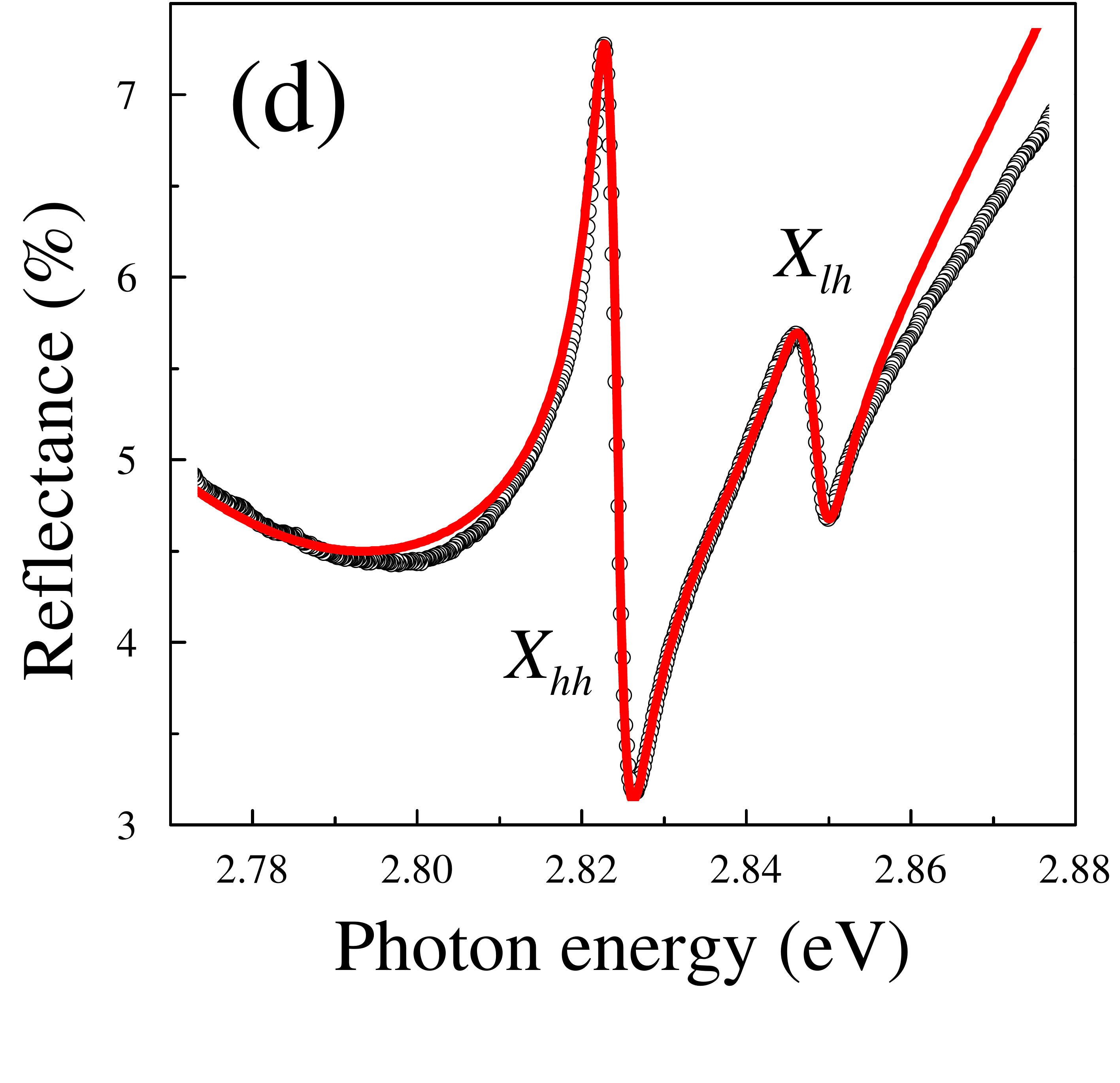}
\caption{(a) Experimental geometry for optical activity registration. The effective magnetic field $\bm B_{\rm eff}$ linear in the light wavevector results in the elliptical polarization of the reflected wave. (b)~Directions of  $\bm B_{\rm eff}$ caused by the bulk inversion asymmetry at various orientations of the light wavevector. (c)~Sample sketch. The widths of the barriers and QW are 
$L=110$~nm, $a=10$~nm.
(d)  Reflectance spectrum of $s$-polarized light incident at angle $\theta=35^\circ$ (symbols) and the fit (solid line). The heavy-hole and light-hole exciton resonances are indicated.
}
\label{fig:sketch}
\end{figure}

Before discussing the experimental results we address
the basic physics of the optical activity and determine requirements
to the experimental geometry.
The optical activity induced polarization conversion can be conveniently described by an effective magnetic field $\bm B_{\rm eff}$ linear in the photon wavevector $\bm q$. $\bm B_{\rm eff}$ affects polarization of the reflected light similarly to a real magnetic field in the magneto-optical Kerr effect.
The effective magnetic field is nonzero due to bulk and structure inversion asymmetries of the QW~\cite{D2dC2v,PRB_2006}.  
Effective magnetic field results in a variety of remarkable effects in exciton physics, mostly studied in double QWs~\cite{LarionovGolub,Butov,Nalitov}.
Optical activity is caused by the part of $\bm B_{\rm eff}$ which has a nonzero projection on $\bm q$, Fig.~\ref{fig:sketch}(a). Therefore structure inversion asymmetry resulting in $\bm B_{\rm eff} \perp \bm q$~\cite{D2dC2v}  does not manifest itself in optical activity.
Bulk and interface inversion asymmetries in the D$_{2d}$ point group result in the effective magnetic field lying in the QW plane. 
Therefore optical activity can be observed only at oblique light incidence.
Direction of $\bm B_{\rm eff}$ depends on the orientation of the photon wavevector in respect to crystallographic axes, Fig.~\ref{fig:sketch}(b). The maximal value of the polarization conversion
is achieved when the incidence plane contains one of cubic axes $\left<100\right>$.

In order to investigate optical activity, we measure light reflection from a QW~\cite{footnote1}. 
This method has been used for investigations of optical activity of gyrotropic bulk semiconductors~\cite{IvchSelk_JETP,rsp_D2d_bulk}. Study of reflection allows detecting the optical activity of QWs in special experimental geometries~\cite{Porubaev}.
In particular, an interaction of the QW with the electric field normal component $E_z$ is necessary. 
As a result, the optical activity is present for the light-hole excitons which have a dipole moment along the growth direction rather than for the heavy-hole excitons which are insensitive to $z$ polarization.
Therefore we choose ZnSe-based QWs where the light-hole exciton is easily observable~\cite{Xlh_1998}.
Polarized spectra of exciton reflection were measured from single QW structures ZnSe/Zn$_{0.82}$Mg$_{0.18}$S$_{0.18}$Se$_{0.82}$.  The samples were grown by molecular beam epitaxy on GaAs epitaxial buffer layers pseudomorphically to GaAs (100) substrates, Fig.~\ref{fig:sketch}(c). 
The growth of ZnMgSSe barriers proceeded at 270~$^\circ$C under the stoichiometric conditions corresponding to equivalent fluxes of the group VI and II elements~\cite{footnote2}.
The total structure thickness including barriers and the QW is 230~nm, which corresponds to $5\lambda/4$ where $\lambda$ is the light wavelength 
at the exciton frequency in ZnSe. 
This allows achieving almost complete 
compensation of reflections from the sample surface and the substrate leading to the pronounced increase of the relative exciton contribution to the reflection~\cite{suppl}.

We studied the dependencies of the reflected light polarization state
on the incidence angle and on orientation of the the incidence plane relative to the crystallographic axes.
The measurements were performed in a glass cylindrical cryostat which allows investigating reflection at arbitrary angles of incidence. 
The sample holder allowed us to rotate the sample around the normal 
by an angle up to 360 degrees.
For measuring reflection spectra, we used a halogen lamp as a light source. The parallel light beam was formed by using lenses and slits. The light spot size exceeded the sample diameter by about two times.  
The light incident on the sample was linearly polarized perpendicular to the plane of incidence ($s$ polarization). All six polarization components of the reflected light were measured.  Namely, two circular intensities $I_{\sigma_\pm}$, two linear ones $I_{s,p}$ that correspond to $s$ and $p$ polarizations, and two linear components in the axes rotated by $\pm 45^\circ$ relative to the plane of incidence, $\tilde{I}_{1,2}$. The spectra were registered by using a 0.5~m  monochromator and a CCD camera.
We estimate the polarization degree measurements accuracy as 0.1~\%.
Polarization state of the reflected light was determined via the Stokes parameters:
\begin{equation}
\label{Stokes}
	P_{circ}={I_{\sigma_+}-I_{\sigma_-} \over I_{\sigma_+}+I_{\sigma_-}}, 
	\quad 
	\tilde{P}_{lin}={\tilde{I}_1-\tilde{I}_2 \over \tilde{I}_1+\tilde{I}_2}. 
\end{equation}
The latter is related to the angle $\alpha$ in Fig.~\ref{fig:sketch}(a) by ${\tilde{P}_{lin}=\sin{2\alpha}}$.

The reflection spectrum at oblique incidence of $s$-polarized light is shown in Fig.~\ref{fig:sketch}(d). Two clearly seen resonances are due to heavy-hole ($X_{hh}$) and light-hole ($X_{lh}$) excitons. 
The spectra do not change qualitatively at variation of the incidence angle $\theta$.
The exciton contribution to the reflectance is big enough owing to a minimum in the background reflection near the exciton frequencies~\cite{suppl}. 
Using the well-established procedure~\cite{EL_book}, we determine the radiative and non-radiative dampings of the light-hole exciton  from the reflection spectrum. They are found to be $\hbar\Gamma_0=0.05$~meV and $\hbar\Gamma=2.35$~meV, respectively.

\begin{figure}[t]
\includegraphics[width=0.47\linewidth]{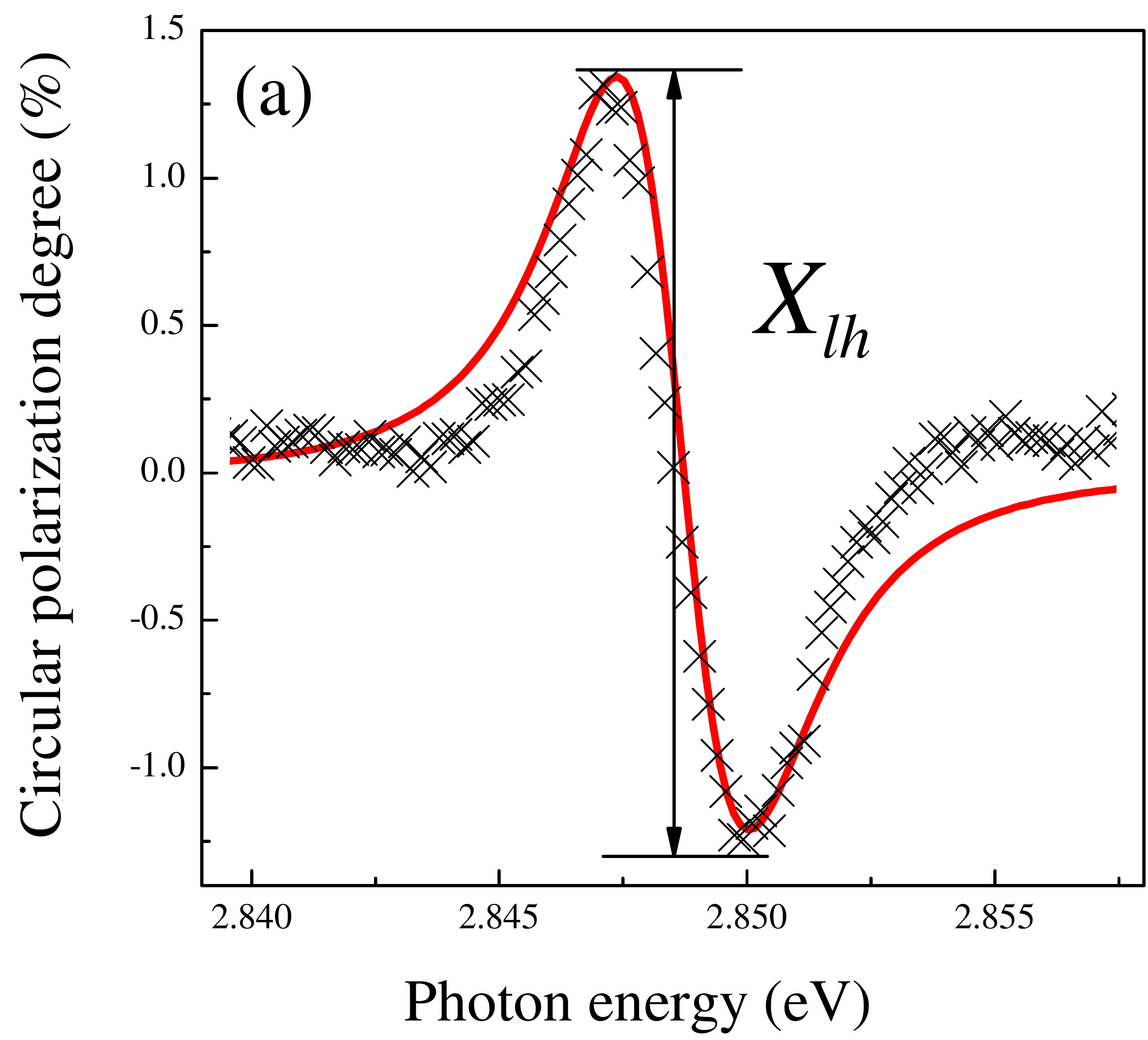}
\quad
\includegraphics[width=0.47\linewidth]{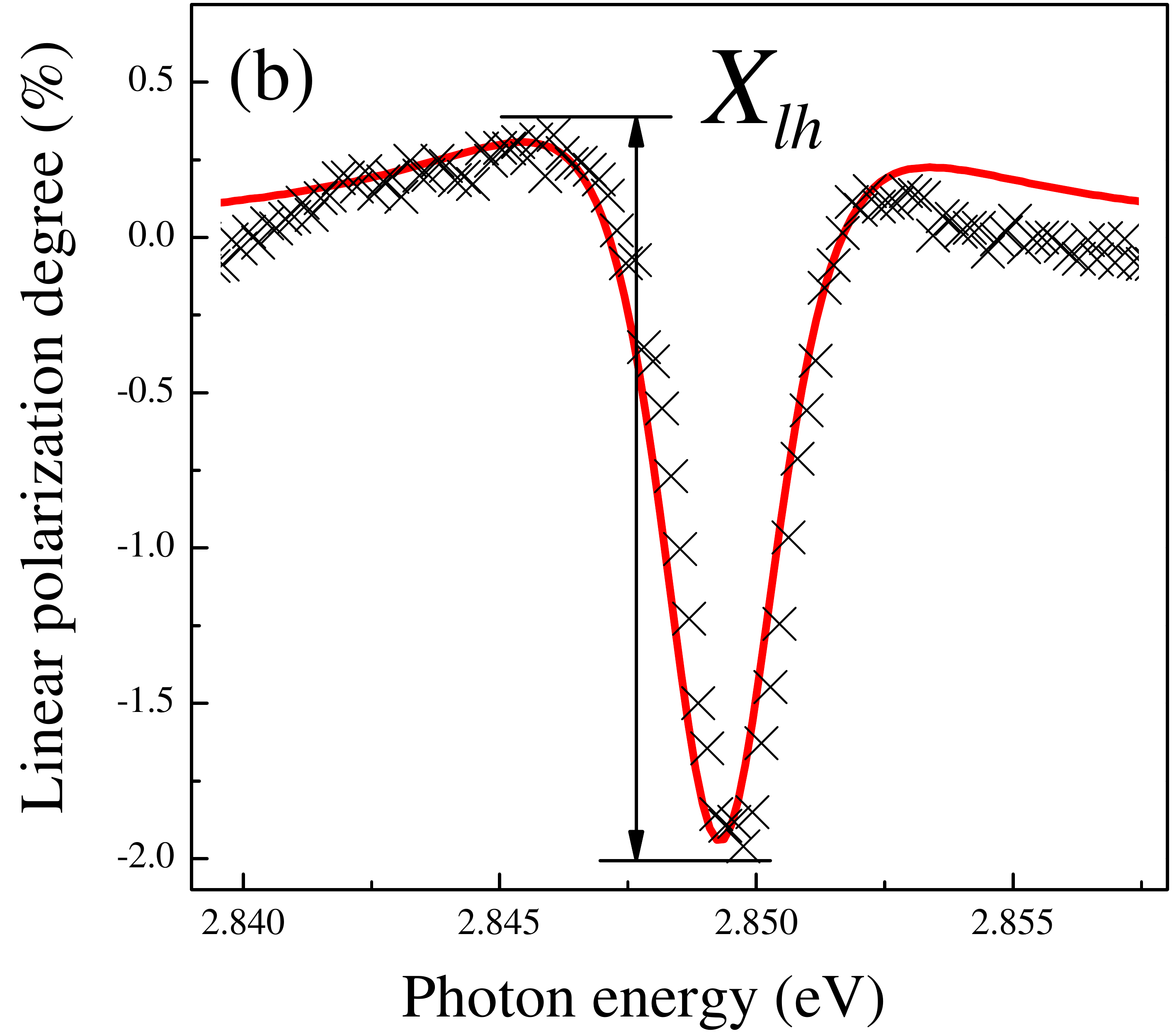}
\\
\includegraphics[width=0.47\linewidth]{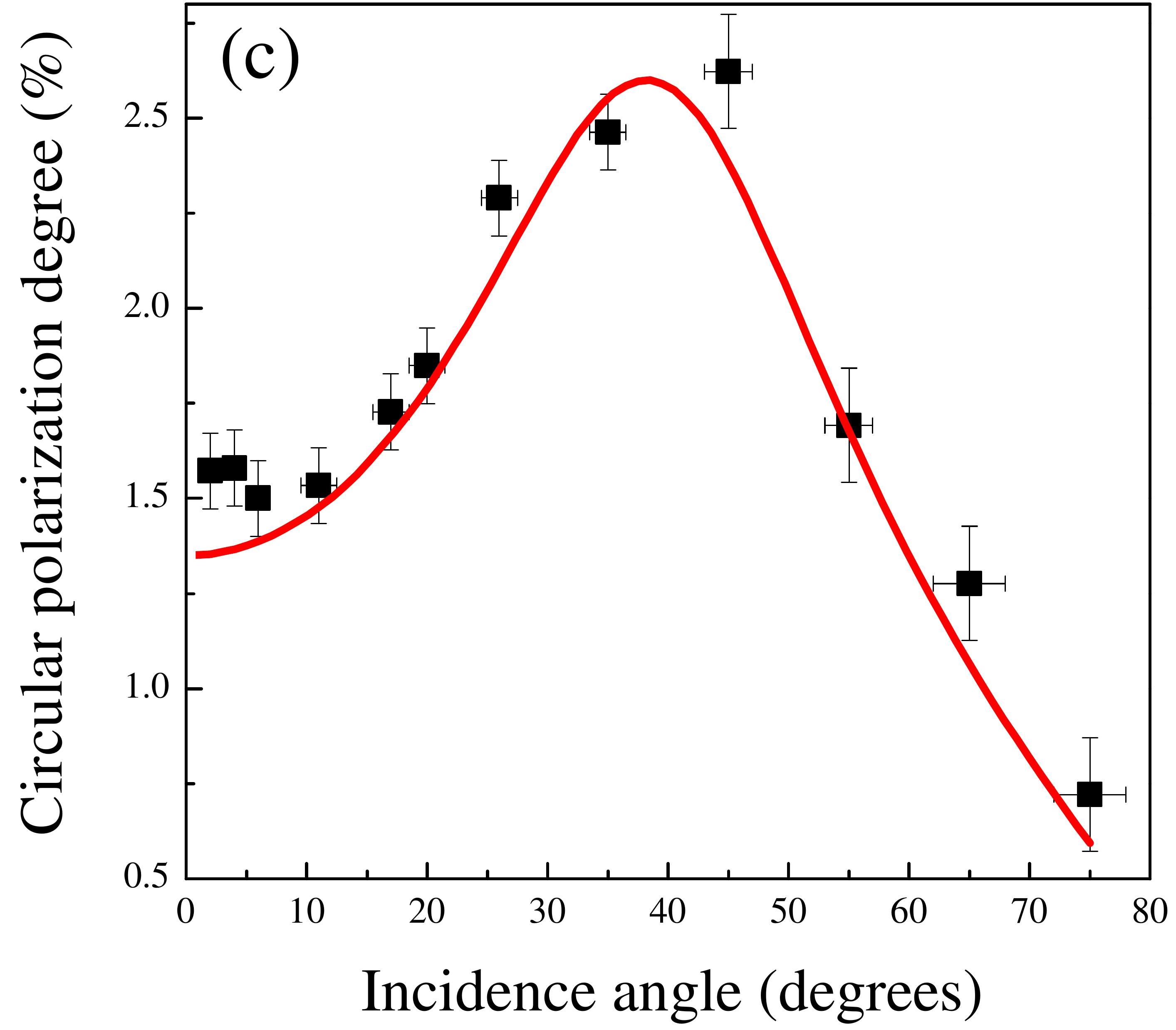}
\quad
\includegraphics[width=0.47\linewidth]{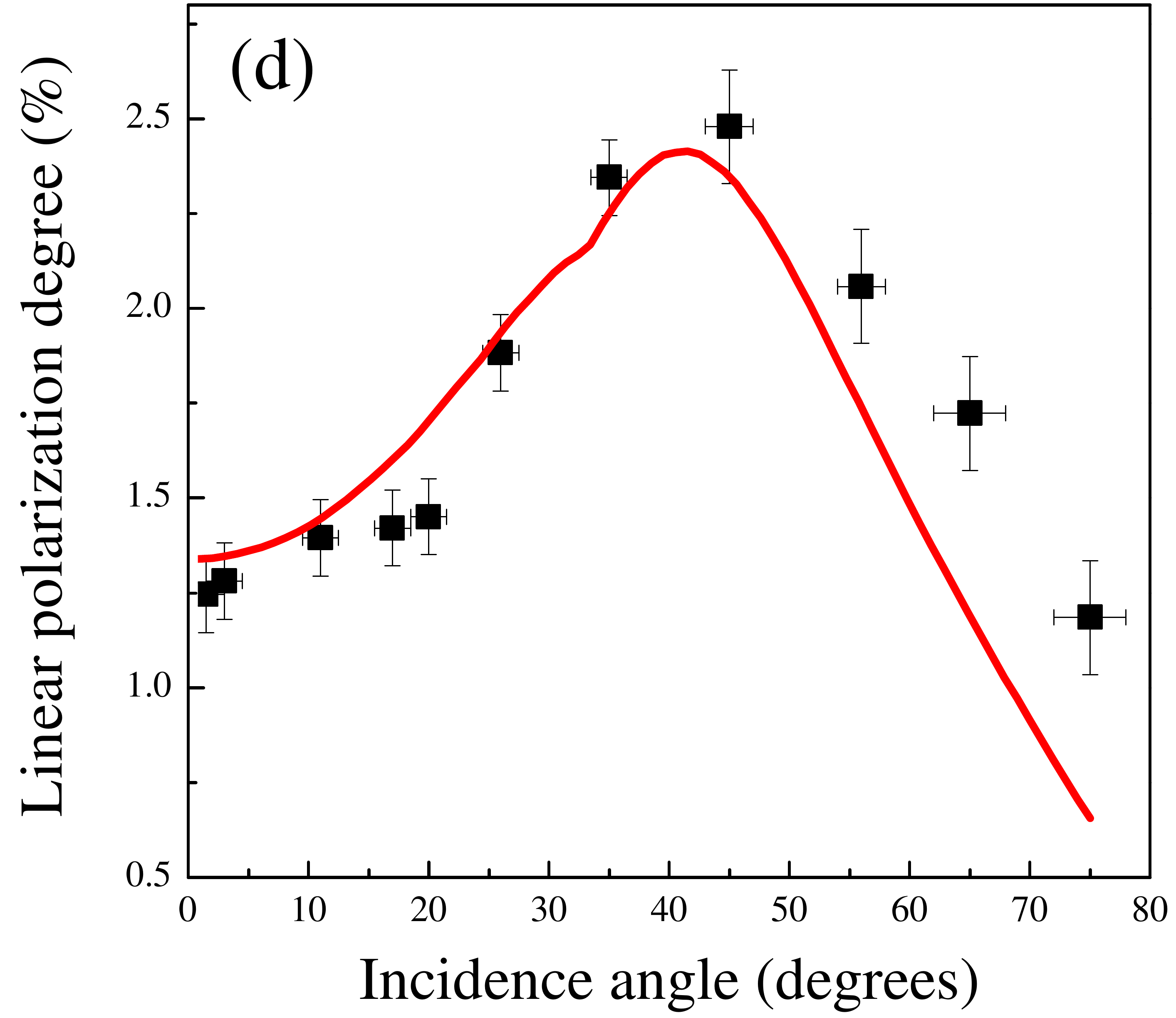}
\caption{Spectra of polarization degrees of reflected light $P_{circ}$~(a) and $\tilde{P}_{lin}$ (b) in the vicinity of the $X_{lh}$ resonance. The wave incident at an angle $\theta = 35^\circ$ is $s$-polarized. Incidence angle dependencies of 
$P_{circ}$ (c) and $\tilde{P}_{lin}$ (d) amplitudes indicated by arrows in panels (a) and (b), respectively.
Solid lines show fit by Eqs.~\eqref{Stokes_rps}.
}
\label{fig:Stokes}
\end{figure}

A presence of the optical activity results in
an appearance of the $p$-polarized component as well as helicity in the reflected wave at incidence of purely $s$-polarized light, Fig.~\ref{fig:sketch}(a). 
Therefore two Stokes parameters
 that are absent in the incident wave, $P_{circ}$ and $\tilde{P}_{lin}$,  are nonzero in the reflected light. 
These two values measured at reflection from our sample are presented in Fig.~\ref{fig:Stokes}. 
Resonant features  at the  $X_{lh}$ frequency are clearly seen in spectral dependencies of $P_{circ}$ and $\tilde{P}_{lin}$, Fig.~\ref{fig:Stokes}(a),(b). Variation of the Stokes parameters with incidence angle are presented in Fig.~\ref{fig:Stokes}(c),(d). The maximal polarization conversion takes place at $\theta \approx 45^\circ$ where it reaches $\approx 2.5$~\%.

\begin{figure}[t]
\includegraphics[width=0.9\linewidth]{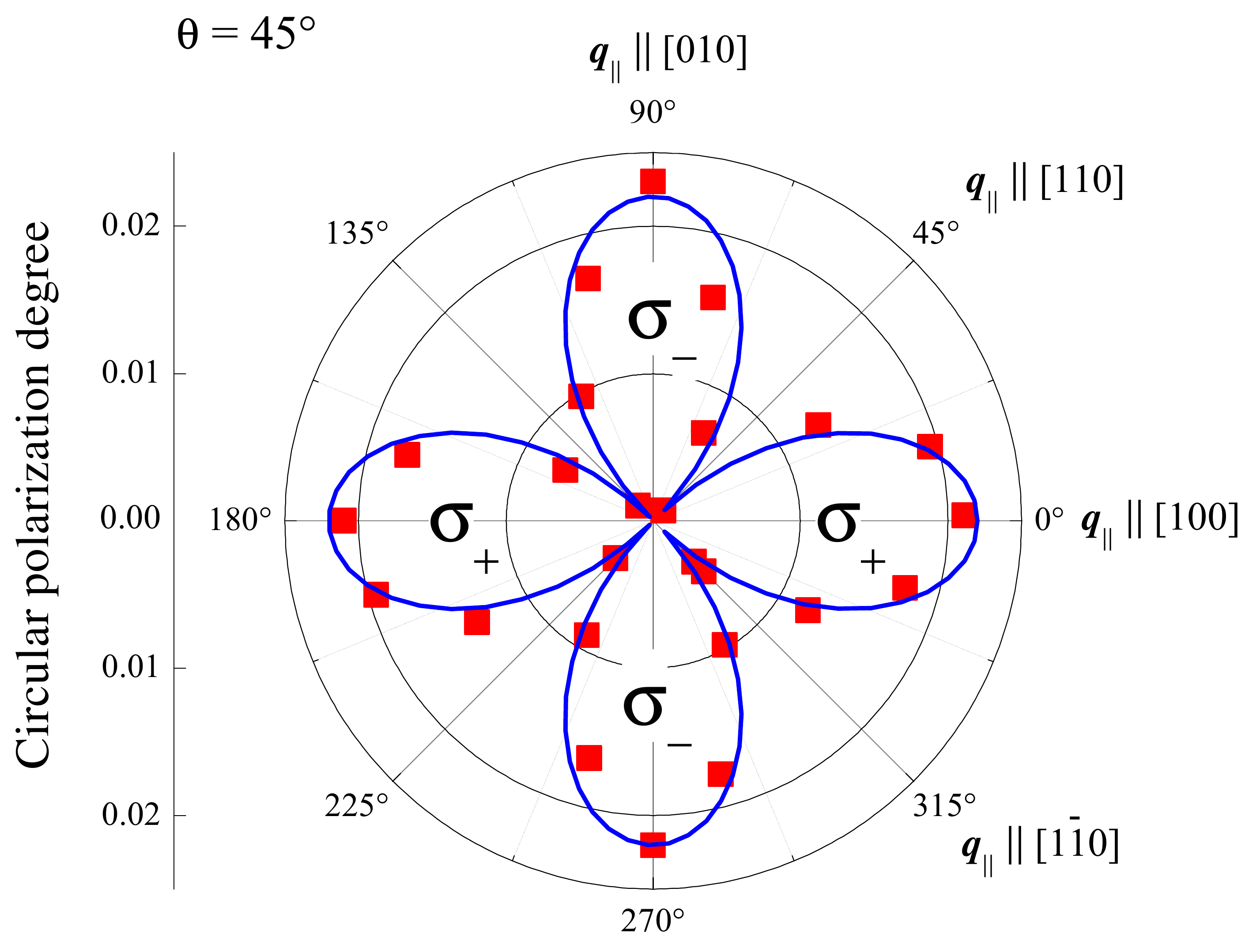}
\caption{Dependence of the circular polarization degree of reflected light on the incidence plane orientation relative to crystallographic axes. Solid line is a fit by $P_{circ}(\varphi)=A\cos{2\varphi}$.}
\label{fig:phi_dep}
\end{figure}

Our measurements show that the Stokes parameters of the reflected light depend on the incidence plane orientation. Figure~\ref{fig:phi_dep} presents the measured dependence $P_{circ}(\varphi)$ where $\varphi$ is an angle between the plane of incidence and the axis [100].
Absolute value of the signal is maximal when the incidence plane contains cubic axes [100], [010]. $P_{circ}$ changes its sign at rotation by $90^\circ$ and reduces to zero  at $\bm q_\parallel$ oriented along $\left< 110\right>$ directions. This behavior reflects the system symmetry and corresponds to the anisotropy of the effective magnetic field $\bm B_\text{eff}$, Fig.~\ref{fig:sketch}(b).

While the explanation of the optical activity effects has
been given in a qualitative way above, we resort now to a
microscopic description based on the equations for the exciton dielectric polarization $\bm P$ in a QW.
Near the light exciton resonance, the microscopic reason for the effective magnetic field resulting in the optical activity is the bulk inversion asymmetry induced spin-orbit interaction. It yields the linear in electron and light hole in-plane momenta $\bm k^{e,h}$ contributions to the single-particle Hamiltonians: 
\begin{equation}
\label{H_i}
	H_i=\beta_i(\sigma_x^i k_x^i - \sigma_y^i k_y^i), \qquad i=e,h,
\end{equation} 
where $\sigma_{x,y}^i$ are the Pauli matrices acting on the spin of the $i^\text{th}$ particle,
$x \parallel [100]$, $y \parallel [010]$ are cubic axes, and  the growth direction is $z \parallel [001]$.
At oblique incidence, the in-plane component of light wavevector is related to $\bm k^{e,h}$ via ${\bm q}_\parallel=\bm k^e + \bm k^h$. 
The term $H_e$ mixes the electron states $S\uparrow$ and $S\downarrow$, and $H_h$ mixes the light hole states 
${\uparrow (X-{\rm i}Y)/\sqrt{6} + \downarrow \sqrt{2/3}Z}$ and ${\downarrow (X+{\rm i}Y)/\sqrt{6} + \uparrow \sqrt{2/3}Z}$, where $S$ is the Bloch orbital in the conduction band, $X,Y,Z$ are the Bloch orbitals in the valence band, and $\uparrow,\downarrow$ are the spinors $\pm 1/2$.
As a result of this mixing, interband transitions are allowed in both in-plane and out-of-plane polarizations, which leads to the polarization conversion, i.e. optical activity.
We stress that the effect is absent for the heavy-hole excitons where the Bloch function has no $Z$ orbital.
The bulk inversion asymmetry results in $\bm q$-linear terms in the equations for the exciton electric polarization~\cite{IvchSelk_JETP,EL_excitons}:
\begin{align}
\label{Pxy}
(\omega^0_{\perp} - \omega) P_{x,y} &- {\rm i}{\beta\over \hbar} \sqrt{d_\perp \over d_\parallel} q_{y,x}
P_z \\
&= d_\perp \Phi(z) \int_{- \infty}^\infty dz' \Phi(z') E_{x,y}(z'), \nonumber \\
\label{Pz}
(\omega^0_{\parallel} - \omega) P_z &+ {\rm i}{\beta\over \hbar} \sqrt{d_\parallel \over d_\perp} (q_xP_y+q_yP_x) \\
&= d_\parallel \Phi(z) \int_{- \infty}^\infty dz' \Phi(z') E_z(z'). \nonumber
\end{align}
Here 
$\bm E$ is the total electric field in the system, the real function $\Phi(z)$ is the wavefunction of the exciton size quantization at coinciding coordinates of electron and hole, 
$\omega^0_{\perp,\parallel}$
and $d_{\perp,\parallel}$
are the frequencies and the squared matrix elements of the light excitons with dipole moments oriented in the QW plane and along $z$, respectively ($d_\parallel / d_\perp=4$). The exciton bulk inversion asymmetry constant 
is related to $\beta_{e,h}$ introduced in Eq.~\eqref{H_i} by 
\begin{equation}
	\beta = {\beta_e m_e + \beta_h m_h \over m_e+m_h}, 
\end{equation}
where $m_{e,h}$ are the electron and the light hole in-plane effective masses.

Solution of the Maxwell equations with the material relations~\eqref{Pxy},~\eqref{Pz} between the polarization and electric field yields
the amplitude $E_p^{QW}$ of the $p$-polarized component reflected from the QW at incidence of $s$-polarized wave with the amplitude $E_{0s}$: ${E_p^{QW} = {\cal R}_{ps}^\beta E_{0s} }$~\cite{suppl} . Here ${\cal R}_{ps}^\beta$ is
the reflection coefficient describing the polarization conversion linear in the spin-orbit exciton constant~$\beta$:
\begin{equation}
\label{Rps}
	{\cal R}_{ps}^\beta = {\sin^2{\theta_1}\over \cos{\theta_1}} \sqrt{d_\parallel \over d_\perp}{ \beta q \cos{2\varphi} \: \Gamma_0 \over (\omega_\perp - \omega - {\rm i}\Gamma)(\omega_\parallel - \omega - {\rm i}\Gamma)},
\end{equation}
where $\theta_1$ is the light propagation angle inside the structure, the radiative and nonradiative $X_{lh}$ linewidths $\Gamma_0$ and $\Gamma$ were determined from the reflection spectrum,
and $\omega_{\perp,\parallel}$
are the light exciton frequencies slightly different from  $\omega^0_{\perp,\parallel}$
due to a radiative renormalization~\cite{EL_book,IK_92}.

Equation~\eqref{Rps} demonstrates that the polarization conversion is absent at normal incidence, and 
the amplitude $E^{QW}_p$ increases as $\theta^2$ at small $\theta$. 
However, the experimental results demonstrate the polarization conversion at normal incidence as well. This effect is not related with the effective magnetic field, but indicates birefringence caused by low symmetry of the real QW under study. One of the reasons for the polarization conversion at normal light incidence may be deformations in the QW plane. 
We describe this effect introducing a mixing of the in-plane components of the exciton polarization 
which does not depend on the wavevector:
\begin{equation}
\label{P_delta}
(\omega^0_\perp - \omega) P_{x,y} + \delta \, P_{y,x}
= d_\perp \Phi(z) \int\limits_{- \infty}^\infty dz' \Phi(z') E_{x,y}(z').
\end{equation}
Here real and imaginary parts of $\delta$ describe, respectively, an energy splitting and  a difference of dampings between the exciton states with dipole moments along $[110]$ and $[1\bar{1}0]$ axes.
Microscopically, the presence of $\delta$ is caused by an effect of in-plane deformations on a short-range exchange interaction  in the exciton.
A finite value of $\delta$ gives rise to the following contribution into the polarization conversion coefficient~\cite{suppl}
\begin{equation}
	{\cal R}_{ps}^\delta = { \delta \cos{2\varphi}\: {\rm i}\Gamma_0 \over (\omega_{\perp} - \omega - {\rm i}\Gamma)^2}.
\end{equation}

The Stokes parameters Eq.~\eqref{Stokes} of the wave reflected from the whole studied structure are described by the complex reflection coefficient $r_{ps}$ relating the amplitudes of the incident  $s$- and reflected $p$-polarized light
as follows~\cite{suppl}
\begin{equation}
\label{Stokes_rps}
	P_{circ}=2{\rm Im}(r_{ps}/r_{ss}), \qquad \tilde{P}_{lin}=2{\rm Re}(r_{ps}/r_{ss}),
\end{equation}
where $r_{ss}$ is the reflection coefficient for $s$-polarized light.
In the studied structure, the resonant signal in the polarization conversion is caused by the QW only.
Therefore $r_{ps}$ is proportional to the reflection coefficient 
describing polarization conversion by 
the QW:
\begin{equation}
	\label{rps}
	r_{ps} = 
	\left( {\cal R}_{ps}^\beta+{\cal R}_{ps}^\delta \right) F(\theta,\omega).
\end{equation}
Here the function $F(\theta,\omega)$ accounts for
multiple reflections from the QW, the sample surface and the interface  with the substrate [Fig.~\ref{fig:sketch}(c)] as well as a conversion of polarization at transmission through the QW~\cite{suppl}.

Comparison of the optical-activity and birefringence coefficients ${\cal R}_{ps}^\beta$ and ${\cal R}_{ps}^\delta$ shows that they have drastically different dependencies on the incidence angle. In contrast to ${\cal R}_{ps}^\beta$ which is zero at normal incidence, ${\cal R}_{ps}^\delta$ is independent of $\theta$. This difference allows us to separate the contributions of optical activity and birefringence into the polarization conversion~\cite{suppl}.
We have fitted both the spectral and incidence-angle dependencies of the Stokes parameters $P_{circ}$ and $\tilde{P}_{lin}$ by Eqs.~\eqref{Stokes_rps},~\eqref{rps}.
Figure~\ref{fig:Stokes} demonstrates that the developed theory describes 
all four dependencies very well.
From the data at normal incidence we determine the birefringence parameter ${\delta=(-0.11\,{\rm i}+0.022)}$~meV.
A larger imaginary value of $\delta$ means that the birefringence of the studied structure is caused 
mainly by a 5~\% difference in the non-radiative dampings $\Gamma$ for the excitons with dipole moments along [110] and $[1\bar{1}0]$ directions rather that in their energy splitting.
The best fit of the data at oblique incidence shown in Fig.~\ref{fig:Stokes} is achieved at the spin-orbit exciton constant $\beta=0.07$~eV\AA.

The $\cos{2\varphi}$ dependence of the Stokes parameters on the angle  $\varphi$ between the polarization plane of incident light and $x$ axis is present in both ${\cal R}_{ps}^\beta$ and ${\cal R}_{ps}^\delta$. This angular dependence perfectly describes the circular polarization degree of the reflected light presented in Fig.~\ref{fig:phi_dep}.

The value of the 
bulk inversion asymmetry 
constant $\beta$ determined from our experiment is in a good agreement with theoretical estimates.
The electron constant $\beta_e$ determined in Ref.~\cite{beta_e} for similar QWs is an order of magnitude smaller than $\beta$ but, as it follows from Refs.~\cite{Zunger,Durnev_etal}, the light-hole spin-orbit splitting exceeds by far the electronic one. The enhancement of $\beta_h$ is most dramatic in the QWs with close ground light-hole level $lh1$ and first excited heavy-hole level $hh2$. In the studied ZnSe-based 10~nm wide QW,  $hh2$ and $lh1$ levels are indeed close to each other.
Therefore we conclude that the  exciton constant is mainly determined by the $lh1$ constant via ${\beta \approx \beta_h m_h/(m_e+m_h)}$, which yields 
$\beta_h \approx 0.14$~eV\AA. 
This value agrees with theoretical estimates~\cite{Zunger,Durnev_etal}.

To summarize, we observed optical activity of semiconductor QWs. The developed theory demonstrates that the polarization conversion is caused by spin-orbit interaction and by birefringence of the studied QW structure. 
The observed effect has a strongly resonant behavior in the vicinity of the light-exciton transition.
Studying the polarization state of reflected light, we determined the exciton spin-orbit spitting in the QW. 

\acknowledgments We thank E.\,L.\,Ivchenko for fruitful discussions and M.\,M.\,Glazov and T.\,V.\,Shubina for helpful comments. The  support from the  Russian Foundation for Basic Research (project 
15-02-04527) is gratefully acknowledged. Theoretical work of L.~E.~G. was supported by Russian Science Foundation (project 14-12-01067). Optical measurements performed by  L.~V.~K. was supported by Russian Science Foundation (project 16-12-10503). The development of MBE technology and sample growth were supported by Russian Science Foundation (project 14-22-00107).

\newpage

\renewcommand{\cite}[1]{{[}\onlinecite{#1}{]}}

\renewcommand{\thepage}{S\arabic{page}}
\renewcommand{\theequation}{S.\arabic{equation}}
\renewcommand{\thefigure}{S\arabic{figure}}
\renewcommand{\bibnumfmt}[1]{[S#1]}
\setcounter{page}{1}
\setcounter{section}{0}
\setcounter{equation}{0}
\setcounter{figure}{0}

\begin{center}
{\bf {ONLINE SUPPLEMENTAL MATERIAL\\
Optical activity of quantum wells}
}
\end{center}

\begin{center}
{\bf S1. Structure design  and experimental details}
\end{center}

\begin{figure}[h]
\includegraphics[width=0.9\linewidth]{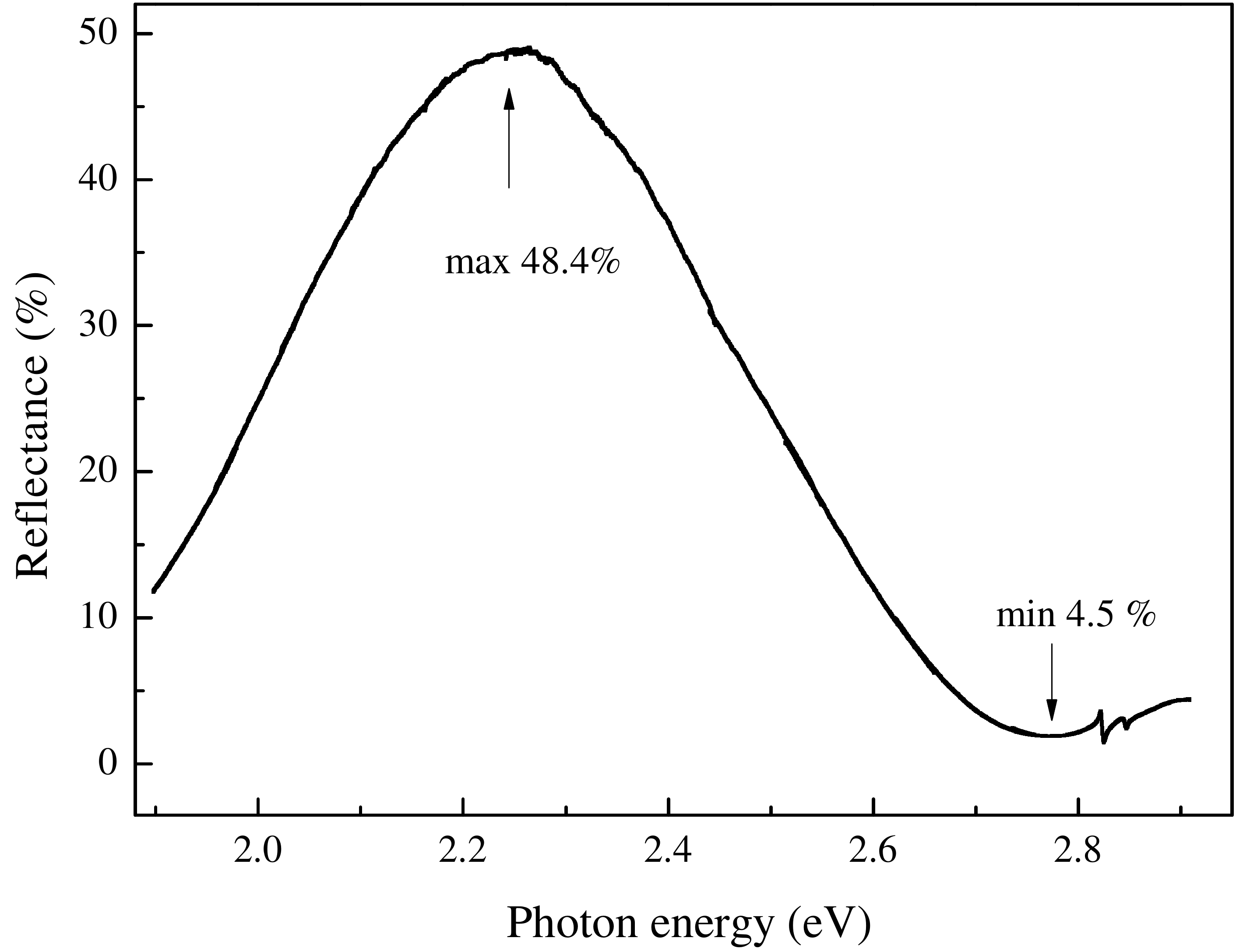}
\caption{Reflectance spectrum in a wide frequency range.}
\label{fig:wide_diap_suppl}
\end{figure}

In order to enhance a role of polarization conversion, we grow a structure of a special design. We choose a ZnSe/ZnMgSSe QW grown on a GaAs substrate. This allowed us to minimize the background reflection not related with the QW. Figure~\ref{fig:wide_diap_suppl} shows the reflection signal in a wide frequency range. The spectrum has both minimum and maximum caused by interference of reflection from the sample surface and the interface with the substrate. Figure~\ref{fig:wide_diap_suppl} demonstrates that the minimal and maximal values of the reflection signal are different by a factor of $\sim 10$, and the light-hole exciton resonance is near the minimum of the background reflection. This allowed us to increase a relative contribution of the QW in the total reflection.

\begin{figure}[t]
\includegraphics[width=0.9\linewidth]{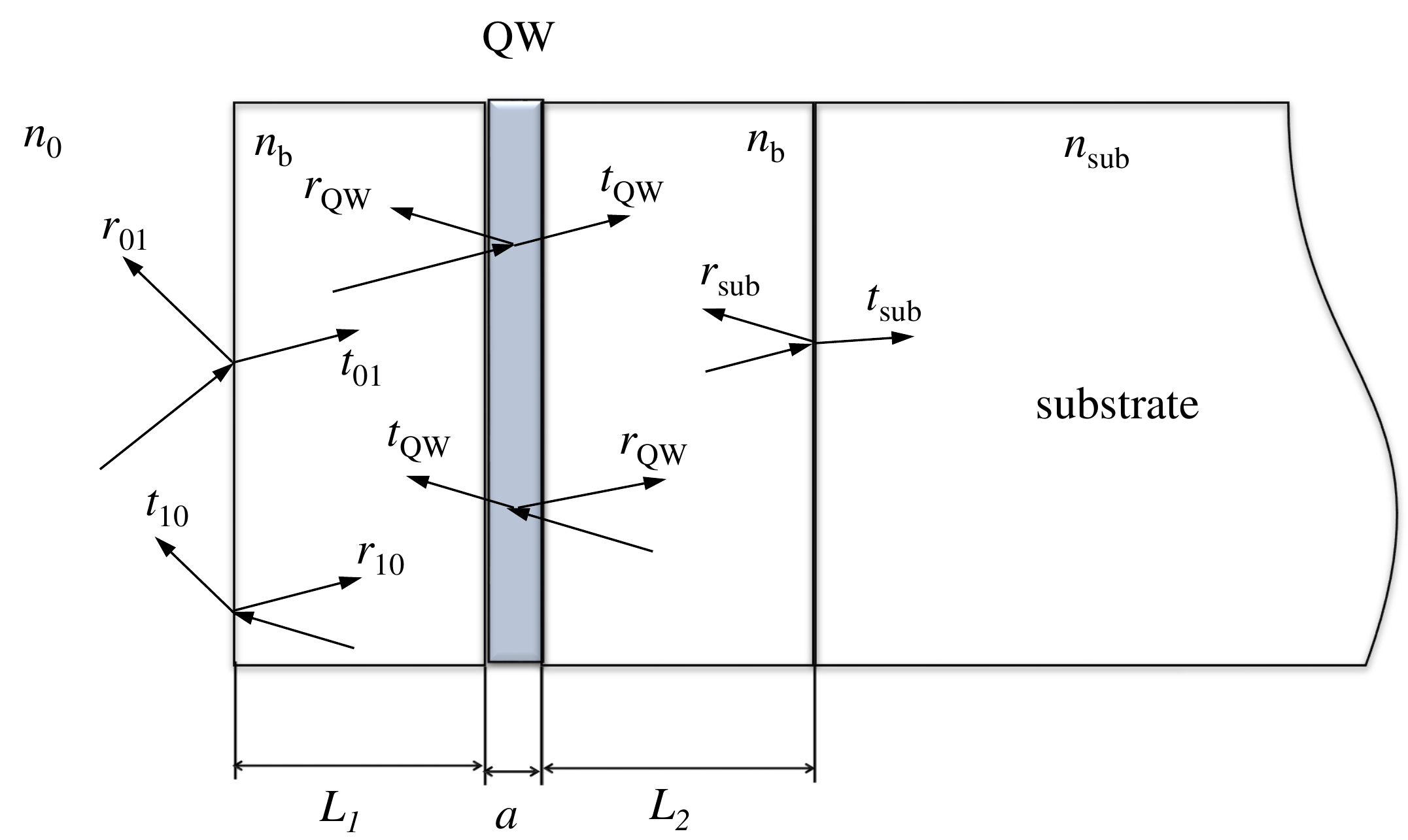}
\caption{Reflection and transmission coefficients in the studied structure.}
\label{fig:refl_suppl}
\end{figure}

In the studied structure, reflection from the substrate is important. Therefore we need to know the refraction index $n_{sub}$ of the GaAs substrate at the exciton frequency of ZnSe. We determine this value from the background reflectance. The maximal and minimal values of the background reflectance are determined by the ratio $r_{01}/r_{sub}$ of the reflection coefficients from the interfaces air/ZnMgSeSe and ZnMgSeSe/GaAs, respectively, Fig.~\ref{fig:refl_suppl}. Neglecting multiple reflections and a dispersion of the refraction index, the ratio of the reflections in the minimum  and in the maximum is given by
\begin{equation}
	\gamma^2 = {R_{min} \over R_{max}} = \left( {r_{01} - r_{sub} \over r_{01} +r_{sub}} \right)^2.
\end{equation}
This yields
\begin{equation}
	{r_{01}  \over r_{sub}} = {1-\gamma \over 1+\gamma}.
\end{equation}
From the experimental value of $\gamma\approx 0.3$, Fig.~\ref{fig:wide_diap_suppl}, and the refraction index $n_b=2.45$~\cite{nb} we obtain $n_{sub} =4.6$. This value is in agreement with the refraction index of GaAs in the studied frequency range~\cite{nsub}.

We note that the sample being placed in the nitrogen vapor was fixed on the heat-conducting copper holder which was partially immersed in liquid nitrogen. This simplifies the analysis of the reflection depndence on the angle of incidence. The sample temperature was controlled by the position of the exciton line in the spectrum at 77~K.

\begin{center}
{\bf S2. Polarization conversion coefficient}
\end{center}

In order to derive the polarization conversion coefficient ${\cal R}_{ps}$ in the vicinity of the light exciton resonance, we find the electric field space distribution from Eqs.~(3),(4) of the main text. 
From Maxwell equations
\begin{equation}
	 {\rm div} \bm D =0, \qquad
	\left( \Delta - {\rm grad} \, {\rm div} \right) \bm E = -\left( {\omega\over c} \right)^2 \bm D,
\end{equation}
and a relation between the electric induction, field and polarization
\begin{equation}
	\bm D = \varepsilon_b \bm E + 4\pi \bm P,
\end{equation}
we get the following equation for $\bm E$ and $\bm P$:
\begin{equation}
	\left( \Delta + q^2 \right) \bm E = - 4\pi  \left({\omega^2\over c^2} + {1\over \varepsilon_b}{\rm grad} \, {\rm div} \right) \bm P.
\end{equation}
Here $\varepsilon_b = n_b^2$ is the background dielectric constant of the barrier and QW materials, and $q=n_b\omega/c$ is the wavevector inside the sample. Hereafter we ignore a possible small difference in the background dielectric constants of QW and barrier materials. This equation can be rewritten in the integral form~\cite{EL_book_suppl}:
\begin{widetext}
\begin{equation}
\label{E}
	\bm E (x,z) = \bm E_0 {\rm e}^{{\rm i}(q_\parallel x + q_z z)} 
	+ {2\pi {\rm i}\over q_z \varepsilon_b} \int\limits_{- \infty}^\infty dz' \, {\rm e}^{{\rm i}q_z |z-z'|} \left(q^2  + {\rm grad} \, {\rm div} \right) \bm P (x,z').
\end{equation}
Here $\bm E_0$ is the amplitude of an incident wave, and we assume that the incidence plane is $(xz)$, $q_z=q\cos{\theta_1}$, ${q_\parallel=q\sin{\theta_1}}$, where $\theta_1$ is the light propagation angle in the barrier material: $\sin{\theta_1} = \sin{\theta}/n_b$.
Substituting the electric field in this form into material Eqs.~(3),(4) of the main text, we come to the closed equation system for the components of the vector $\bm P$. The coordinate dependence is $\bm P(x,z) = \bm C \Phi(z) {\rm e}^{{\rm i}q_\parallel x}$ with a constant vector $\bm C$. Here $\Phi(z)=\Phi(-z)$ is the wavefunction of the exciton size quantization at coinciding coordinates of electron and hole ($\Phi$ is chosen to be real). 
Components of the vector $\bm C$ satisfy the following linear equation system:
\begin{align}
	(\omega_p - \omega - {\rm i}\Gamma_p)C_x - {\rm i}{\beta q_\parallel\over \hbar}\sin{2\varphi} \sqrt{d_\perp \over d_\parallel} C_z = d_\perp \Lambda_0 E_{0p}\cos{\theta_1}, \\
	(\omega_s - \omega - {\rm i}\Gamma_s)C_y - {\rm i}{\beta q_\parallel\over \hbar}\cos{2\varphi} \sqrt{d_\perp \over d_\parallel} C_z = d_\perp \Lambda_0 E_{0s}, \\
	(\omega_\parallel - \omega - {\rm i}\Gamma_\parallel)C_z + {\rm i}{\beta q_\parallel\over \hbar}(C_x\sin{2\varphi}+C_y\cos{2\varphi}) \sqrt{d_\parallel \over d_\perp} = -d_\parallel \Lambda_0 E_{0p}\sin{\theta_1}.
	\end{align}
Here $\varphi$ is the angle between the plane of incidence and the axis [100], $\Lambda_0=\int\limits_{- \infty}^\infty dz \, \Phi(z) \cos{q_z z}$, $\Gamma_{p,s}=\Gamma_{p,s}^0+ \Gamma_\perp$, $\Gamma_\parallel=\Gamma_\parallel^0+ \Gamma_z$ with  $\Gamma_{\perp,z}$ being the dampings caused by nonradiative processes, and
\begin{equation}
	\omega_{p,s}-{\rm i}\Gamma_{p,s}^0 = \omega_\perp^0 + (\delta\omega-{\rm i}\Gamma_0) (\cos{\theta_1})^{\pm 1},
	\qquad
	\omega_\parallel- {\rm i}\Gamma_\parallel^0 = \omega_\parallel^0 + \Delta + (\delta\omega-{\rm i}\Gamma_0) {d_\parallel \over d_\perp} {\sin^2{\theta_1} \over \cos{\theta_1}},
\end{equation}
with~\cite{EL_book_suppl}
\begin{equation}
\Gamma_0 = {2\pi q d_\perp \over \varepsilon_b}\Lambda_0^2, 
\qquad
\delta\omega = {2\pi q d_\perp \over \varepsilon_b}\int\limits_{-\infty}^\infty dz \int\limits_{-\infty}^\infty dz' \Phi(z)\Phi(z') \sin(q_z|z-z'|),
\qquad
	\Delta = {4\pi d_\parallel \over \varepsilon_b}\int\limits_{-\infty}^\infty \Phi^2(z) dz.
\end{equation}

Finding the vector $\bm C$, we get from Eq.~\eqref{E} the electric field in the whole space. At $z \to -\infty$, it has a form 
\begin{equation}
	\bm E (x,z \to -\infty) = \left(
	\begin{array}{c}
		E_{0s} \\
		E_{0p}
	\end{array}
	 \right) {\rm e}^{{\rm i}(q_\parallel x + q_z z)} + 
	\left(
	\begin{array}{cc}
		r_s^{QW} & {\cal R}_{ps}\\
		{\cal R}_{ps} & r_p^{QW}
	\end{array}
	 \right)
	\left(
	\begin{array}{c}
		E_{0s} \\
		E_{0p}
	\end{array}
	 \right) {\rm e}^{{\rm i}(q_\parallel x - q_z z)} ,
\end{equation}
and in the limit  $z \to \infty$ we have:
\begin{equation}
	\bm E (x,z \to \infty) = \left(
	\begin{array}{c}
		E_{0s} \\
		E_{0p}
	\end{array}
	 \right) {\rm e}^{{\rm i}(q_\parallel x + q_z z)} + 
	\left(
	\begin{array}{cc}
		t_s^{QW} & {\cal T}_{ps}\\
		{\cal T}_{ps} & t_p^{QW}
	\end{array}
	 \right)
	\left(
	\begin{array}{c}
		E_{0s} \\
		E_{0p}
	\end{array}
	 \right) {\rm e}^{{\rm i}(q_\parallel x + q_z z)} .
\end{equation}
\end{widetext}
Here the reflection and transmission coefficients for $s$ and $p$ polarized light incident on the QW are given by
\begin{equation}
\label{r_QW}
	r_s^{QW}= {{\rm i}\Gamma^0_{s}\over {\omega}_{s}-\omega - {\rm i}\Gamma_{s}},
	\quad
	t_s^{QW}=1+r_s^{QW},
\end{equation}
\begin{align}
	r_p^{QW} = {{\rm i}\Gamma^0_{p}\over {\omega}_{p}-\omega - {\rm i}\Gamma_{p}}-
		 {{\rm i}\Gamma^0_\parallel \over {\omega}_{\parallel}-\omega - {\rm i}\Gamma_\parallel},\\
	t_p^{QW} = 1+{{\rm i}\Gamma^0_{p}\over {\omega}_{p}-\omega - {\rm i}\Gamma_{p}}+
		 {{\rm i}\Gamma^0_\parallel\over {\omega}_{\parallel}-\omega - {\rm i}\Gamma_\parallel}. \nonumber
	\end{align}

The reflection and transmission coefficients describing the polarization conversion are related by
\begin{equation}
\label{Tps}
	{\cal T}_{ps} = - {\cal R}_{ps}.
\end{equation}
In the first order in $\beta$, the polarization conversion reflection coefficient is given by
\begin{equation}
\label{Rps_suppl}
	{\cal R}_{ps}^\beta = {\sin^2{\theta_1}\over \cos{\theta_1}} \sqrt{d_\parallel \over d_\perp}{ \beta q \cos{2\varphi} \: \Gamma_0 \over (\omega_s - \omega - {\rm i}\Gamma_s)(\omega_\parallel - \omega - {\rm i}\Gamma_\parallel)}.
\end{equation}

A nonequivalence of [110] and $[1\bar{1}0]$ directions in the QW plane results in birefringence which leads to the polarization conversion  even for the normal incidence. Microscopically, we describe this effect by a difference $\pm \delta$ between the complex eigenfrequencies for excitons with dipole moments oriented along these directions, see Eq.~(7) of the main text. For a light wave incident in the $(xz)$ plane, this leads to a coupled equations for the coefficients $C_x$ and $C_y$:
\begin{align}
\label{C_delta}
(\omega_{p} - \omega- {\rm i}\Gamma_p + \delta \sin{2\varphi} ) C_x +& \delta \cos{2\varphi} C_y \nonumber \\
&= d_\perp \Lambda_0 E_{0p}\cos{\theta_1}, \\
(\omega_{s} - \omega- {\rm i}\Gamma_s - \delta \sin{2\varphi} ) C_y + &\delta \cos{2\varphi} C_x
= d_\perp \Lambda_0 E_{0s}. \nonumber
\end{align}
Solution of this system linear in $\delta$ yields the conversion coefficient in the following form:
\begin{equation}
\label{Rps_delta}
	{\cal R}_{ps}^\delta = { \delta \cos{2\varphi}\: {\rm i}\Gamma_0 \over (\omega_s - \omega - {\rm i}\Gamma_s)(\omega_p - \omega - {\rm i}\Gamma_p)}.
\end{equation}

Taking in Eqs.~\eqref{Rps_suppl} and~\eqref{Rps_delta} $\omega_s \approx \omega_p = \omega_\perp$, $\Gamma_s \approx \Gamma_p \approx \Gamma_\parallel = \Gamma$, we obtain  Eqs.~(6) and~(8) of the main text.

\begin{center}
{\bf S3. Reflection and transmission coefficients for the QW structure}
\end{center}

The Stokes parameters of the reflected light are given by
\begin{equation}
		P_{circ}= \text{i} {E_sE_p^* - E_s^*E_p \over |E|^2}, \quad
		\tilde{P}_{lin}={E_sE_p^* + E_s^*E_p \over |E|^2},
\end{equation}
where $E_{s,p}$ are the components of the reflected wave. At incidence of $s$-polarized light with an amplitude $E_{0s}$, the reflected field components are $E_s=r_{ss}E_{0s}$, $E_p=r_{ps}E_{0s}$ with $|r_{ps}| \ll |r_{ss}|$. Therefore we have:
\begin{equation}
\label{Stokes_suppl}
	P_{circ}=2{\rm Im}(r_{ps}/r_{ss}), \qquad \tilde{P}_{lin}=2{\rm Re}(r_{ps}/r_{ss}).
\end{equation}

The coefficient $r_{ss}$ is the reflection coefficient from the whole structure, Fig.~\ref{fig:refl_suppl}, for $s$-polarized light:
\begin{equation}
\label{suppl_rss}
	r_{ss} =  r_{01}^s + {t_{01}^s t_{10}^s {\cal R}_s {\rm e}^{2{\rm i}\varphi_1} \over 1+ r_{01}^s {\cal R}_s {\rm e}^{2{\rm i}\varphi_1}},
	\end{equation}
where
	\begin{equation}	
	{\cal R}_s = r_{QW}^s + { (t_{QW}^s)^2 r_{sub}^s {\rm e}^{2{\rm i}\varphi_2} \over 1-r_{QW}^s r_{sub}^s {\rm e}^{2{\rm i}\varphi_2}}.
\end{equation}
Here $\varphi_{1,2}=q(L_{1,2}+a/2)\cos{\theta_1}$ are the phases for traveling from the interface barrier/air (barrier/substrate) to the QW center, and
we introduce the Fresnel reflection and transmission coefficients at the interface between air and the structure,
see Fig.~\ref{fig:refl_suppl}. Namely, for oblique incidence from air, we denote them as $r_{01}^i$, $t_{01}^i$, and at oblique incidence from the structure as $r_{10}^i$, $t_{10}^i$ ($i=s,p$). 
The reflection coefficients at incidence from the barrier material on the substrate are denoted as $r_{sub}^{i}$.

\begin{widetext}
\begin{figure*}[t]
\includegraphics[width=0.8\linewidth]{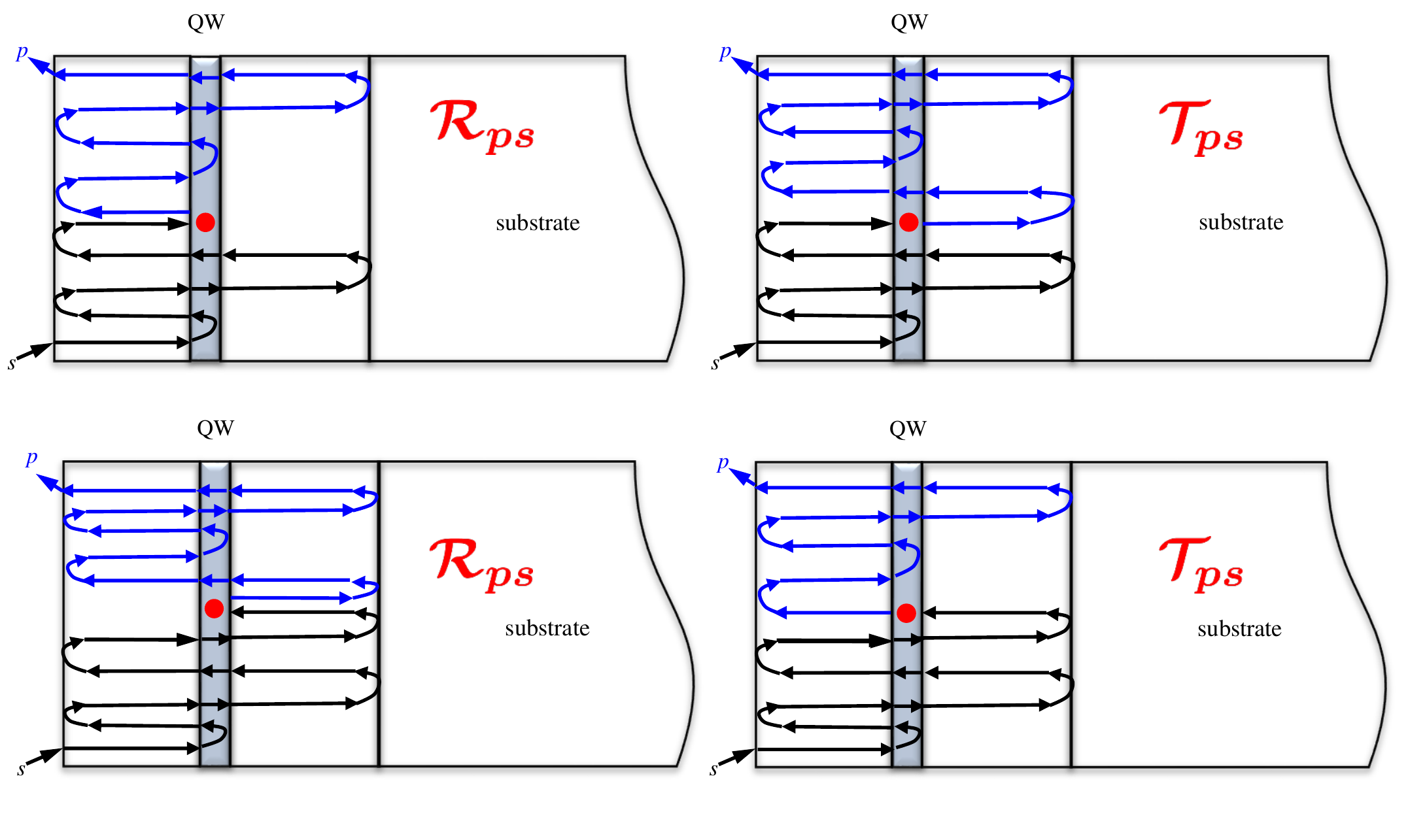}
\caption{Light paths in the studied structure. The red circle shows the moment of the polarization conversion.}
\label{fig:four_proc}
\end{figure*}
\end{widetext}

The polarization conversion reflection coefficient is a sum of four contributions shown in Fig.~\ref{fig:four_proc}:
\begin{align}
\label{rps_suppl}
	r_{ps} =  t_{10}^p \Psi_1^p \Psi_1^s t_{01}^s 
		(1+G_s+Q_s)(1+G_p+Q_p) \\
		\times
			[{\cal R}_{ps}+{\cal T}_{ps}t_{QW}^s (\Psi_2^s)^2r_{sub}^s+t_{QW}^p (\Psi_2^p)^2r_{sub}^p {\cal T}_{ps} \nonumber \\
			+  t_{QW}^p (\Psi_2^p)^2r_{sub}^p{\cal R}_{ps} t_{QW}^s (\Psi_2^s)^2r_{sub}^s]. 
	\nonumber 
\end{align}
Here multiple reflections from the QW  and at the interfaces with air and with substrate (Fig.~\ref{fig:refl_suppl}) are taken into account as well as a conversion of polarization at transmission through the QW.
We introduce (for $i=s,p$) the reflection coefficient being a sum of amplitudes for passing a ``short''  path (from the sample surface to the QW and back)
and a ``long'' path (from the sample surface to the substrate and back):
\begin{equation}
	G_i = r_{10}^i(\Psi_1^i)^2 \left[r_{QW}^i+(t_{QW}^i\Psi_2^i)^2r_{sub}^i \right],
\end{equation}
and the amplitude for passing both ``short'' and ``long'' paths without polarization conversion:
\begin{equation}
	Q_i = r_{QW}^ir_{sub}^i \left[r_{10}^it_{QW}^i(\Psi_1^i)^2\Psi_2^i \right]^2.
\end{equation}
The factors $\Psi_{1,2}^i$ describe multiple transmissions of light between the QW and the interface with air as well as between the QW and the substrate:
\begin{equation}
\Psi_1^i = {{\rm e}^{{\rm i}\varphi_1} \over 1-r_{10}^i r_{QW}^i {\rm e}^{2{\rm i}\varphi_1}},
	\quad
		\Psi_2^i = {{\rm e}^{{\rm i}\varphi_2} \over 1-r_{sub}^i r_{QW}^i {\rm e}^{2{\rm i}\varphi_2}}. 
\end{equation}

The four terms in square brackets in Eq.~\eqref{rps_suppl} account for the $s \to p$ polarization conversion 
at reflection and at transmission of light incident on the QW from the left and from the right,
see Fig.~\ref{fig:four_proc}. 
Taking into account the relation~\eqref{Tps}, we obtain from Eq.~\eqref{rps_suppl} Eq.~(3) of the main text:
\begin{equation}
	r_{ps} = {\cal R}_{ps} \, F(\theta,\omega)
\end{equation}
with
\begin{align}
F(\theta,\omega)&	 = t_{10}^p \Psi_1^p \Psi_1^s t_{01}^s 
		(1+G_s+Q_s)(1+G_p+Q_p) \nonumber  \\
		&\times
		[1-t_{QW}^s (\Psi_2^s)^2r_{sub}^s][1-t_{QW}^p (\Psi_2^p)^2r_{sub}^p]. 
\end{align}

\begin{center}
{\bf S4. Optical activity and birefringence contributions}
\end{center}

In order to separate the optical activity and the birefringence contributions to the measured signal we used their different dependence on the incidence angle. Equations~(6) and~(8) of the main text show that ${\cal R}_{ps}^\delta$ is independent of $\theta$ while ${\cal R}_{ps}^\beta$ monotonously increases with $\theta$. However, the measured values of the Stokes parameters are also determined by the amplitude reflection coefficient $r_{ss}$, see Eqs.~\eqref{Stokes_suppl}, which also depends on $\theta$. 
Despite the reflection coefficient of $s$ polarized light from a single surface increases monotonously with $\theta$, the angular dependence in our structure is more complicated due to the discussed above interference of reflections from the surface and the substrate.
\begin{figure}[h]
\includegraphics[width=0.8\linewidth]{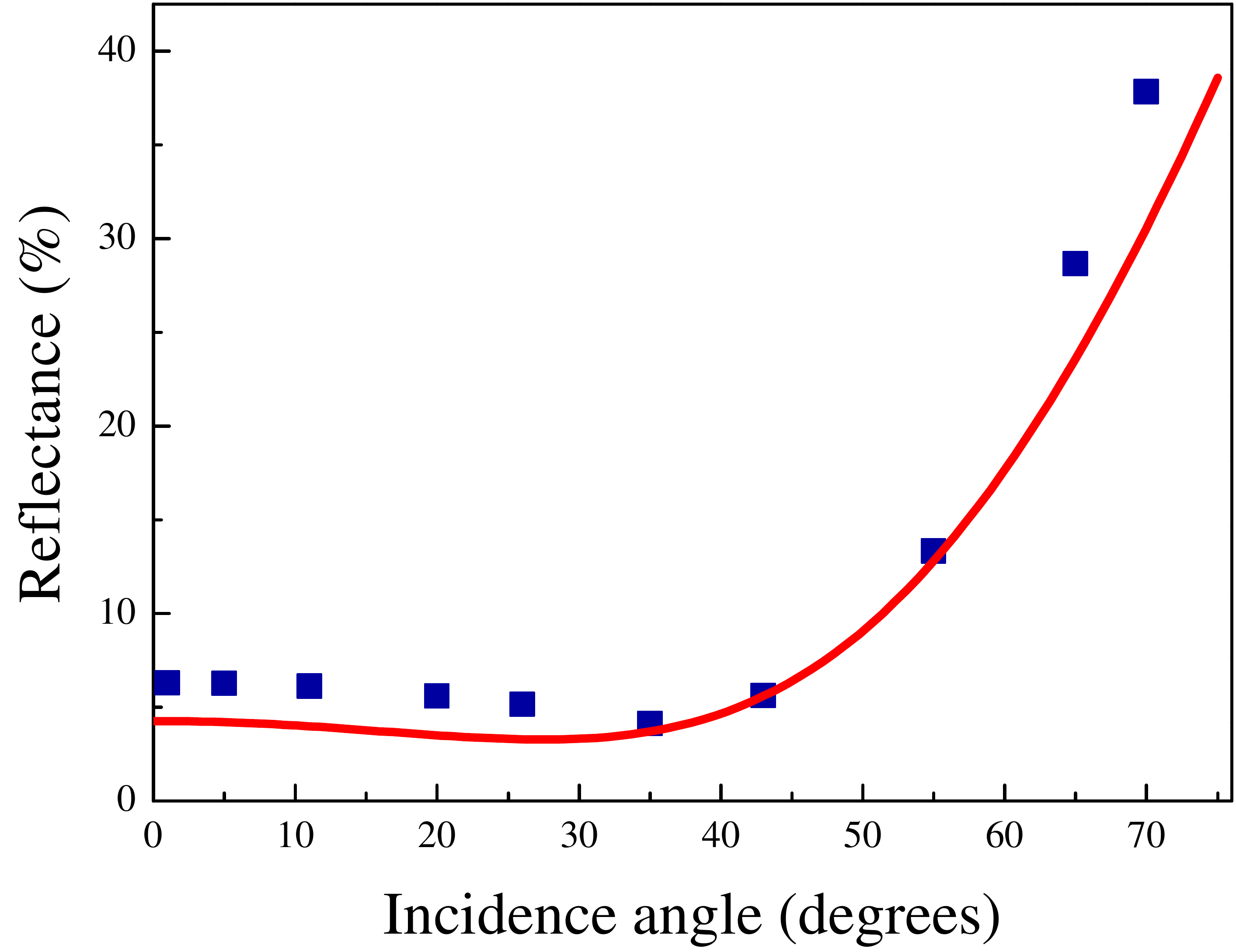}
\caption{Dependence of the background reflectance in the vicinity of $X_{lh}$ resonance on the incidence angle. Solid line is the fit by Eq.~\eqref{suppl_rss}.}
\label{fig:rss_suppl}
\end{figure}
Figure~\ref{fig:rss_suppl} shows the background reflectance dependence on $\theta$ near the light-hole exciton resonance. It demonstrates that $r_{ss}$ is almost constant for $\theta <35^\circ$, and it raises monotonously for higher incidence angles. Therefore, the increase of the Stokes parameters at $\theta \approx 35^\circ \ldots 45^\circ$ seen in Figs.~2(c),(d) of the main text is caused solely by the contribution of optical activity.


\begin{thebibliography}{99}

\bibitem{Zvezdin_Kotov} A. K. Zvezdin and V. A. Kotov, \textit{Modern Magnetooptics and Magnetooptical Materials}, 
(Institute of Physics Publishing, Bristol and Philadelphia, 1997).

\bibitem{metamat} 
N. K. Grady, J. E. Heyes, D. R. Chowdhury, Y. Zeng, M. T. Reiten, A. K. Azad, A.J. Taylor, D. A. R. Dalvit, and H.-T. Chen, 
Science \textbf{340}, 1304 (2013).

\bibitem{metamat1} 
Mengxin Ren,	Eric Plum,	Jingjun Xu, and Nikolay I. Zheludev, Nature Commun. \textbf{3}, 833 (2012).


\bibitem{swast} 
S. V. Lobanov, S. G. Tikhodeev, N. A. Gippius, A. A. Maksimov, E. V. Filatov, I. I. Tartakovskii, V. D. Kulakovskii, T. Weiss, C. Schneider, J. Ge{\ss}ler, M. Kamp, and S. H\"ofling, Phys. Rev. B \textbf{92}, 205309 (2015).

\bibitem{twist} 	
X. M. Xi, T. Weiss, G. K. L. Wong, F. Biancalana, S. M. Barnett, M. J. Padgett, and P. S. J. Russell,
Phys. Rev. Lett. \textbf{110}, 143903 (2013).

\bibitem{OSHE} 	
C. Leyder, M. Romanelli, J. Ph. Karr, E. Giacobino, T. C. H. Liew, M. M. Glazov, A. V. Kavokin, G. Malpuech and A. Bramati, Nat. Phys. \textbf{3}, 628  (2007).


\bibitem{birefr_g-fact} 
Yu. G. Kusrayev, A. V. Koudinov, I. G. Aksyanov, B. P. Zakharchenya, T. Wojtowicz, G. Karczewski, and J. Kossut, Phys. Rev. Lett. \textbf{82}, 3176 (1999).

\bibitem{birefr_local_fields} 
A. A. Sirenko, P. Etchegoin, A. Fainstein, K. Eberl, and M. Cardona, phys. stat. sol. (b) \textbf{215}, 241 (1999).

\bibitem{birefr_deform} A.V. Koudinov, N.S. Averkiev, Yu.G. Kusrayev, B.R. Namozov, B.P. Zakharchenya,
D. Wolverson, J.J. Davies, T. Wojtowicz, G. Karczewski, J. Kossut,
Phys. Rev. B \textbf{74}, 195338 (2006).

\bibitem{Loginov} 
D. K. Loginov, P. S. Grigoryev, Yu. P. Efimov, S. A. Eliseev, V. A. Lovtcius, V. V. Petrov,
E. V. Ubyivovk, and I. V. Ignatiev, Phys. Stat. Sol. B  (2016), doi: 10.1002/pssb.201552735.


\bibitem{birefr_QWs_Moldova} 
N. Syrbu, A. Dorogan, V. Dorogan, V. Zalamai, Superlatt. Microstr. \textbf{82}, 143 (2015).


\bibitem{birefr_dots} 
T. Kiessling, A. V. Platonov, G. V. Astakhov, T. Slobodskyy, S. Mahapatra, W. Ossau, G. Schmidt, K. Brunner,
and L. W. Molenkamp, Phys. Rev. B \textbf{74}, 041301(R) (2006).

\bibitem{birefr_pol_conv_dots} 
G. V. Astakhov, T. Kiessling, A. V. Platonov, T. Slobodskyy, S. Mahapatra, W. Ossau, G. Schmidt, K. Brunner, and L. W. Molenkamp, Phys. Rev. Lett. \textbf{96}, 027402 (2006).

\bibitem{IIA_Voisin} 
O. Krebs, D. Rondi, J. L. Gentner, L. Goldstein, and P. Voisin,
Phys. Rev. Lett. \textbf{80}, 5770 (1998).

\bibitem{IIA_doped} 
J. L. Yu, Y. H. Chen, X. Bo, C. Y. Jiang, X. L. Ye, S. J. Wu, and H. S. Gao,
J. Appl. Phys. \textbf{113}, 083504 (2013).


\bibitem{Top1}
W. Gao, M. Lawrence, B. Yang, F. Liu, F. Fang, B. B\'{e}ri, J. Li, and S. Zhang,
Phys. Rev. Lett. \textbf{114}, 037402 (2015).

\bibitem{Top2}  
S. Zhong, J. Orenstein, and J. E. Moore,
Phys. Rev. Lett. \textbf{115}, 117403 (2015).


\bibitem{AgrGinzb} V. M. Agranovich and V. L. Ginzburg, \textit{Crystal Optics
with Spatial Dispersion, and Excitons} (Springer-Verlag, Berlin, 1984).


\bibitem{IvchSelk_JETP} 
E.L. Ivchenko, A.V. Sel'kin, Sov. Phys. JETP \textbf{49}, 933 (1979).

\bibitem{bulk_gyr} P. Etchegoin and M. Cardona, Solid State Commun. \textbf{82}, 655 (1992).


\bibitem{BeTeZnSe} 
D. R. Yakovlev, E. L. Ivchenko, V. P. Kochereshko, A. V. Platonov, S. V. Zaitsev, A. A. Maksimov, I. I. Tartakovskii, V. D. Kulakovskii, W. Ossau, M. Keim, A. Waag, and G. Landwehr
Phys. Rev. B \textbf{61}, 2421(R) (2000).

\bibitem{D2dC2v} 
 S. D. Ganichev and L. E. Golub, Phys. Status Solidi B \textbf{251}, 1801 (2014).


\bibitem{Golub_EPL} 	
L. E. Golub, EPL \textbf{98}, 54005 (2012). 

\bibitem{Porubaev} 
L. E. Golub and F. V. Porubaev, Phys. Solid State \textbf{55}, 2239 (2013).

\bibitem{PRB_2006} 
N.S. Averkiev, L.E. Golub, A.S. Gurevich, V.P. Evtikhiev, V.P. Kochereshko, A.V. Platonov, A.S. Shkolnik, Yu.P. Efimov,  Phys. Rev. B \textbf{74}, 033305 (2006).



\bibitem{LarionovGolub} 
A.V. Larionov and L.E. Golub, Phys. Rev. B \textbf{78}, 033302 (2008).

\bibitem{Butov} A. A. High, A. T. Hammack, J. R. Leonard, S. Yang, L. V. Butov, T. Ostatnick\'y, M. Vladimirova, A. V. Kavokin, T. C. H. Liew, K. L. Campman, and A. C. Gossard,
Phys. Rev. Lett. \textbf{110}, 246403 (2013).

\bibitem{Nalitov} 
A. V. Nalitov, D. D. Solnyshkov, N. A. Gippius, and G. Malpuech, Phys. Rev. B \textbf{90}, 235304 (2014).


\bibitem{footnote1} A possible reason why the optical activity has not been observed in QWs so far is that typically it is detected by the rotation of the light polarization as it propagates through a  medium. Measurements of this kind are hardly realizable in real QWs due to a small transmission through a substrate.

\bibitem{rsp_D2d_bulk} 
O. Arteaga, Opt. Lett. \textbf{40}, 4277 (2015).

\bibitem{Xlh_1998} 
A. V. Platonov, V. P. Kochereshko, D. R. Yakovlev, U. Zehnder, W. Ossau, W. Faschinger, and G. Landwehr,
Phys. Solid State  \textbf{40},  745 (1998).


\bibitem{footnote2} Special four-period ZnSSe/MgSe superlattices with the same average composition as the bulk Zn$_{0.82}$Mg$_{0.18}$S$_{0.18}$Se$_{0.82}$ barrier and the total thickness of 10~nm each were grown at both interfaces of the ZnSe QW to improve the interface flatness.


\bibitem{suppl}
See Supplemental Material for details on the experimental setup, sample growth, and theoretical description of  the polarization conversion.

\bibitem{EL_book} E. L. Ivchenko, \textit{Optical Spectroscopy of Semiconductor Nanostructures} (Alpha Science Int., Harrow, UK, 2005).


\bibitem{EL_excitons} 
E.L. Ivchenko in \textit{Excitons}, E.I. Rashba and M.D. Sturge (eds.), Vol. 2 of \textit{Modern Problems in Condensed Matter Sciences} (North-Holland, Amsterdam, New York, 1982).

\bibitem{IK_92} E. L. Ivchenko, A. V. Kavokin, Sov. Phys. Solid State  \textbf{34}, 968 (1992).

\bibitem{beta_e} 
H. Mino, S. Yonaiyama, K. Ohto, and R. Akimoto, Appl. Phys. Lett. \textbf{99}, 161901 (2011).

\bibitem{Zunger} 
 J.-W. Luo, A. N. Chantis, M. van Schilfgaarde, G. Bester, and
A. Zunger, Phys. Rev. Lett. \textbf{104}, 066405 (2010).

\bibitem{Durnev_etal} 
M. V. Durnev, M. M. Glazov, and E. L. Ivchenko, Phys. Rev. B \textbf{89}, 075430 (2014).

\end{thebibliography}

\begin{thebibliography}{3}
\bibitem{nb} 
M. Ukita, H. Okuyama, M. Ozawa, A. Ishibashi, K. Akimoto and Y. Mori, Appl. Phys. Lett. \textbf{63}, 2082 (1993).

\bibitem{nsub} 
H. R. Philipp and H. Ehrenreich, Phys. Rev. \textbf{129}, 1550 (1963).

\bibitem{EL_book_suppl} E. L. Ivchenko, \textit{Optical Spectroscopy of Semiconductor Nanostructures} (Alpha Science Int., Harrow, UK, 2005).

\end{thebibliography}
\end{document}